\documentstyle[12pt,psfig]{article}
\newcommand{\mbf}[1]{\vec{#1}}
\newcommand{\ie}{{\em i.e.}}                            %    i.e.
\newcommand{\de}{\Delta}
\begin{document}
%==========================================================================
\begin{titlepage}

{\small \sl Physics Letters B}\hfill ECT$^*$-97-003, LBNL-40207\\[12ex]

\begin{center}{\large {\bf
Effects of Spin-Isospin Modes in Transport Simulations$^\dagger$}}\\[8ex]
{\sl Johan Helgesson$^a$ and J\o rgen Randrup$^b$}\\[2ex]

$^a$ECT$^*$, European Centre for Studies in Theoretical Nuclear Physics 
                      and Related Areas, Trento, Italy\\[1ex]

$^b$Nuclear Science Division, 
         Lawrence Berkeley National Laboratory,\\
         Berkeley, California 94720\\[6ex]
April 23, 1997\\[6ex]
{\sl Abstract:}\\
\end{center}

{\small\noindent
In-medium properties derived for nuclear matter
in a microscopic $\pi + N N^{-1} + \Delta N^{-1}$ model 
are incorporated into transport simulations of nuclear collisions
by means of a local-density approximation
and by utilizing a local medium frame.
Certain features of the transport results differ from those
based on the corresponding vacuum properties.
Comparisons of the $\pi$ and $\Delta$ production rates, 
as well as pion energy spectra, are discussed in particular.

\vfill
\noindent
{\sl PACS:}
%	25.70.-z,	%	Low and intermediate energy heavy-ion reactions
	25.75.Dw,	%	Particle and resonance production
	25.75.-q,	%	Relativistic heavy-ion reactions
	24.10.Cn,	%	Many-body theory
	13.75.Gx	%	Pion-baryon interactions
\\

\\

\noindent
$^\dagger$This work was supported 
by the Training and Mobility through Research (TMR) programme 
of the European Community under contract ERBFMBICT950086 
and by the Director, Office of Energy Research,
Office of High Energy and Nuclear Physics,
Nuclear Physics Division of the U.S. Department of Energy
under Contract No.\ DE-AC03-76SF00098.\\
}%      end small
\end{titlepage}

% =======================================================
\section{Introduction}
% =======================================================

Collisions between heavy nuclei at intermediate energies
produce hot and dense hadronic matter
that may be probed by means of the energetic particles emitted
\cite{Metag,Cassing,Mosel}.
The collision dynamics have been fairly well understood
within microscopic transport models, 
such as BUU and QMD \cite{Cassing,Mosel,Wolf,Aichelin},
in which the hadrons are propagated in an effective one-body field
while subject to direct elastic and inelastic two-body collisions.
In particular,
sufficiently energetic $NN$ collisions may agitate
one or both of the collision partners to a nucleon resonance.
Such resonances propagate in their own mean field
and may collide with nucleons or other nucleon resonances as well.
Furthermore, the nucleon resonances may decay by meson emission
and these decay processes constitute the main mechanisms
for the production of energetic mesons \cite{Mosel}.

Normally,
the transport descriptions employ
the vacuum properties of the baryon resonances and mesons,
\ie\ the cross sections, decay widths, and dispersion relations are taken
as their (measured or inferred) values in vacuum \cite{Wolf}.
However, as is well known,
the strong interaction between pions, nucleons, and $\Delta$ isobars 
may generate spin-isospin modes in nuclear matter.
While most of these modes are non-collective in their character,
being dominated by a single baryon-hole excitation,
others are collective and correspond to meson-like states (quasi-mesons)
that may be important in the transport description.

Various incorporations of such in-medium modifications 
to transport simulations of nuclear collisions have already been made
on the basis of a simple two-level $\Delta N^{-1}$-model
\cite{Weise,Bertsch,Giessen,Texas,Tub}.
A more consistent set of in-medium quantities,
suitable for implementation in transport descriptions,
was derived in a refined $\pi + N N^{-1} + \Delta N^{-1}$ model \cite{simAP}.
These in-medium properties, 
and their implementation into transport models,
were thoroughly discussed in refs.\ \cite{simAP,simNP},
but no explicit transport simulation has yet been performed.

This work present first results
from transport simulations including in-medium effects
from the refined $\pi + N N^{-1} + \Delta N^{-1}$ model of ref.\ \cite{simAP}.
We aim to gain a qualitative impression of the degree to which
the in-medium properties in idealized nuclear matter
survive the transport simulations and lead to observable effects.
To keep matters simple,
we include only the most important properties of the spin-isospin modes,
deferring a more rigorous treatment for later,
and we consider only the $\pi$-like spin-longitudinal collective modes, 
since they are dominant at the energies considered
(the $\rho$-like spin-transverse collective modes can be treated analogously).
 
% =====================================================
\newpage
\section{Spin-isospin modes in matter}
\label{sec_SIMinNM}
% =====================================================

Relative to the simple two-level $\de N^{-1}$ model
utilized in previous transport treatments
\cite{Weise,Bertsch,Giessen,Texas,Tub},
some features change significantly with the more
rigorous treatment of the $\de N^{-1}$ model which 
includes bands of $\de N^{-1}$ and $N N^{-1}$ states.
The most pertinent points are briefly recalled below
(a detailed discussion was given in ref.\ \cite{simAP}).

Dispersion relations of spin-isospin modes are calculated
in a non-relativistic RPA formalism,
treating interactions between pion, nucleon-hole, and $\Delta$-hole states.
In such a model one should in principle 
treat the $\Delta$ width self-consistently,
{\em i.e.} the $\Delta$ isobars that are used as components 
in the spin-isospin modes 
should contain the in-medium $\Delta$ width. 
However, such a treatment would encompasses processes like 
$
   \tilde{\pi}_j    \rightarrow    \Delta N^{-1} 
                    \rightarrow    (N+\tilde{\pi}_k) N^{-1}
$
which are already included explicitly in the transport simulations
by processes like
$
   \tilde{\pi}_j + N   \rightarrow    \Delta  
                       \rightarrow    N+\tilde{\pi}_k 
$.
We have therefore omitted the $\Delta$ width
in the present microscopic treatment 
(see ref.\ \cite{simAP} for a self-consistent treatment of the $\Delta$ width).
However, we want to emphasize that this choice is not unambiguous, 
since the omission of the $\Delta$ width
implies the omission of certain other effects, 
{\em e.g.} that the pionic modes
have a Breit-Wigner-like distribution of energies, 
just as the $\Delta$ resonance
has a Breit-Wigner like distribution of invariant masses.

Furthermore it was found in ref.\ \cite{simAP} 
that the dependence on the temperature is rather weak 
up to moderately large temperatures. 
Although incorporation of in-medium effects depending on $T>0$ 
into transport models is rather straightforward, 
the treatment in the transport formalism will depend on one 
more variable which makes it more cumbersome.
Therefore, in the present microscopical calculations 
we only use properties calculated for matter at zero temperature.

The in-medium properties are obtained by 
using the Green's function technique,
starting from non-interacting hadrons.
Standard $p$-wave interactions \cite{simAP,OTW}
are used at the
$N \pi N$ and $N \pi \Delta$ vertices 
and effective short-range interactions at
baryon-hole vertices.
The strength of the short-range interactions 
is determined by the correlation parameters 
$g_{NN}'$, $g_{N\Delta}'$, and $g_{\Delta \Delta}'$.

The spin-isospin modes are obtained within the RPA approximation, 
symbolically
\begin{eqnarray}
\label{eq_GRPA}
&~&\	G^{\rm RPA}(\alpha,\beta;\omega)\
	=\	G_{0}(\alpha,\beta;\omega)\\ \nonumber
&~&\	+\ \sum_{\gamma,\kappa} G_{0}(\alpha,\gamma;\omega)\
                           {\cal V}(\gamma,\kappa;\omega)\
	G^{\rm RPA}(\kappa,\beta;\omega)\ .
\end{eqnarray}
The spin-isospin modes, 
here represented by the Green's function 
$G^{\rm RPA}$,
are in this approximation 
obtained as an infinite iteration of
(non-interacting) pion, nucleon-hole, 
and $\Delta$-hole states,
represented by the diagonal Green's function $G_0$, 
coupled by the symbolic interaction ${\cal V}$.
A set of RPA equations equivalent to eq.\ (\ref{eq_GRPA}) 
were derived in ref.\ \cite{simAP}
and they determine the eigenvectors and eigenenergies 
for the different spin-isospin modes.
The eigenvectors yield the amplitudes of the different components
($\pi$, $NN^{-1}$, $\Delta N^{-1}$)
forming the particular spin-isospin eigenmode with the given energy.
These RPA amplitudes contain important information 
about the nature of the different spin-isospin modes. 

\begin{figure}	%.............................................................
\centerline{\psfig{file=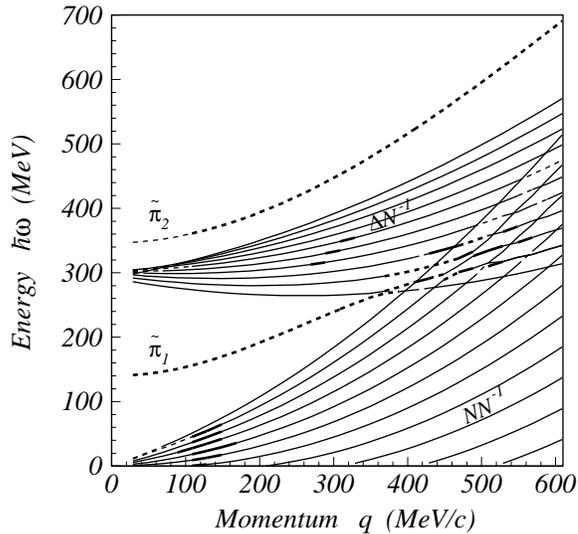,height=8.0cm,angle=0}}
\caption{The spin-longitudinal spin-isospin modes in nuclear matter
         at normal density and zero temperature,
         as obtained with the $\pi + N N^{-1} + \Delta N^{-1}$ model.
         The solid curves represent non-collective modes, while
         dashed curves are used to indicate collective strength.
         Thick lines indicate that the pionic component 
         is larger than 5\%.}
\label{fig_DispT0}
\end{figure}

The dispersion relation\footnote
  {The dispersion relation is obtained in a finite box 
   with periodic boundary conditions, giving a finite number of
   $NN^{-1}$ and $\Delta N^{-1}$ states. Infinite nuclear matter
   is approached as the size of the box is increased.}
is shown in Fig.\ \ref{fig_DispT0}
for normal nuclear density, $\rho$=$\rho_0$,
and at zero temperature, $T$=0.
There are two collective modes,
corresponding to those of the simple two-level model.
They are often referred to as the pion 
and $\Delta N^{-1}$ branch, respectively.
However in their structure these modes are similar and 
by the collectivity they are both pion-like 
in the sense that many such modes may be created 
(in contrast to a non-collective mode that is exhausted by a single mode). 
Therefore the two collective spin-isospin modes 
should be treated on an equal footing, 
and they can effectively be regarded as particles 
of mesonic character 
(quasi-mesons, here denoted $\tilde{\pi}_1$ and $\tilde{\pi}_2$).
In addition,
a number of  $N N^{-1}$ and $\Delta N^{-1}$ like modes are obtained.
These modes are mainly non-collective,
each being dominated by a single $N N^{-1}$
or $\Delta N^{-1}$ component.
	%.............................................................

The total $\Delta$ width gives the transition rate
for the $\Delta$ resonance to decay to any of its decay channels.
In a transport description one explicitly allows 
the $\Delta$ resonance to decay 
into specific final particles.
Consequently, one needs not only the total $\Delta$ width
(which is the sum of all decay channels)
but also the partial widths governing the decay 
into specific RPA channels.
These decay channels consist
of a nucleon and one of the spin-isospin modes.
The partial $\Delta$ width for a $\Delta$ decay to a nucleon
and a spin-longitudinal mode $\nu$ becomes 
\begin{eqnarray}
           \Gamma^\nu_\Delta(E_\Delta,\mbf{p}_\Delta) 
& = &
           \frac{1}{3} \int \frac{ d^3 q }{ (2 \pi)^3 }
           \bar{n}(\mbf{p}_\Delta-\mbf{q})
           |h^\nu_{N\Delta}|^2 \nonumber
\\ 
& \times &
           2 \pi \: \delta(E_\Delta - e_N(\mbf{p}_\Delta-\mbf{q}) 
                         - \hbar \omega_\nu)\ ,
\label{eq_GnuReal}
\end{eqnarray}
where $e_N$ is the energy of the nucleon.
The factor $h^\nu_{N\Delta}$ is obtained 
from the interactions at the vertex
consisting the $\Delta$, the nucleon,
and the spin-isospin mode $\nu$.
The interactions to be used depend 
on the non-interacting states
that the mode consists of 
and must therefore be multiplied
by the amplitude of the corresponding state
(for further details, see ref.\ \cite{simAP}).
Note that when this expression is to be used in transport models
the Pauli blocking factor $\bar{n}_N$ should be omitted 
since the Pauli blocking is treated explicitly 
in the transport description. 

The total $\Delta$ width is obtained as a sum over 
all modes, collective as well as non-collective, 
\begin{equation}
  \Gamma_\Delta(E_\de,\mbf{p}_\Delta) 
= 
  \sum_\nu \Gamma^\nu_\Delta(E_\Delta,\mbf{p}_\Delta)\ .
\end{equation}
(The sum contains also spin-transverse contributions \cite{simAP}
which are not treated explicitly here.)

Unfortunately, the available experimental data is insufficient
to fully constrain the model parameters and so some choices must be made.
A particular set of values was motivated in ref.\ \cite{simAP}.
It contains a density-dependent effective nucleon mass
and reproduces the vacuum cross sections $p+p \rightarrow \Delta^{++}+n$  
as well the imaginary part of the $\Delta$-nucleus spreading potential 
of ref.\ \cite{Hirata}. 
In this work we incorporate the calculated medium effects 
into an established transport code that utilizes vacuum properties
\cite{LiBauer}. 
Since that code does not admit effective masses,
we have taken $m_N^* = m_N$
 and then readjusted $g^\prime_{\Delta \Delta}$ 
(from 0.35 to 0.5) to reproduce the same data,
keeping all other parameters remaining as in ref.\ \cite{simAP}. 

% ====================================================================
\section{Incorporation of the in-medium properties}
\label{sec_SIMtoTM}
% ====================================================================
In this section we discuss how the results 
from the microscopical calculations in infinite nuclear matter
is incorporated in a dynamical transport simulation 
of a heavy ion collision.
For this purpose we will compare with 
a simple test-particle description 
that propagates nucleons ($N$), delta isobars ($\Delta$), 
and pions ($\pi$) \cite{LiBauer}. 
This will be referred to as the {\em standard} transport description. 
In this standard description, 
the properties of the $\Delta$'s and $\pi$'s 
(decay widths, cross sections, dispersion relations, {\em etc.}) 
are taken to be those in vacuum. 
In ref.\ \cite{simAP} it was thoroughly discussed 
how these vacuum properties can be replaced 
by in-medium properties in a consistent way.
Here we present a brief recapitulation 
of the points essential for this study.

The general idea is to employ a local-density approximation
in order to incorporate the infinite-matter properties
obtained by the microscopic calculations
into the transport treatment of the nuclear collision dynamics.
Furthermore, the microscopic calculations are performed in a system 
where the medium is at rest
and the obtained dispersion relations and decay widths 
refer to this {\em medium frame}. 
When incorporating the in-medium quantities into the transport formalism, 
this requirement can be met by employing a local medium frame
in which the current density vanishes.

In ref.\ \cite{simAP} it 
was found that
at all densities 
the spin-isospin modes 
(in the spin-longitudinal channel) 
could be well categorized as 
either non-collective $\Delta N^{-1}$ modes, 
non-collective $N N^{-1}$ modes,
or one of two different collective modes corresponding to quasi-pions. 
The non-collective $N N^{-1}$ and $\Delta N^{-1}$ modes 
correspond to particles and processes 
already incorporated in standard transport models.
On the other hand,
the collective modes reflect the modified properties of the pion in the medium
and they should therefore replace the real pions
present in the standard transport treatment. 
Both correspond to quasi-pions,
each propagating with their own Hamiltonian.
In accordance with the replacement of the real pions 
with two different types of quasi-pions, 
appropriate partial $\Delta$ decay widths must be used 
(as will will be further discussed below).

The quasi-pions do not only have a pion component, 
but also $N N^{-1}$ and $\Delta N^{-1}$ components. 
The strength of these components 
vary with the momentum of the pionic mode 
and with the nuclear density, 
and this variation is different 
for the two types of pionic mode.
Fortunately their specific structure 
is irrelevant\footnote{ For secondary processes, 
                        such as dilepton production 
                        from pion annihilation, 
                        the amplitudes become important also in the medium
			(see, for example, ref.\ \cite{HR:ee}). }, 
as as long as these quasi-particles remain well inside 
the nuclear medium. 
When the surrounding density approaches zero
(as happens if the quasi-pion escapes through the nuclear surface
or as the result of an overall expansion),
one of the components of the quasi-pions will automatically be realized.
That is to say, it will turn into either 
a free pion or an unperturbed $\Delta N^{-1}$ state. 
The number of quasi-pions that are realized 
as $\Delta N^{-1}$ states is quite small,
as will be discussed in section \ref{sec_Res}. 
Below a certain critical density,
these few modes could be converted to a free $\Delta$ 
by absorption on a nearby nucleon, 
but in the present study we simply simply let 
also these quasi-pions be realized as free pions.

The pionic modes are created in the $\Delta$ decays 
and governed by the $\Delta$ decay width.
While the $\Delta$ in vacuum
has only the single decay channel $\Delta \rightarrow N + \pi$, 
there are several channels available in the medium.
Apart from the decay to the quasi-pions 
there is a contribution from the decay channels
$ \de \rightarrow N + N N^{-1}$ 
and
$ \de \rightarrow N + \Delta N^{-1}$.
However these decay channels are already included 
in the standard transport treatment by the processes
$ \Delta + N \rightarrow N + N$
and
$ \Delta + N \rightarrow \Delta + N$,
and these partial widths should therefore be omitted.

The in-medium partial decay width 
differs from the free width for mainly two reasons. 
First, the phase space for the decay is different, 
{\em i.e.} the pionic mode has a different energy-momentum relation, 
$\hbar \tilde{\omega}(q)$. 
Second, the pion component of the quasi-pion is no longer unity, 
as it has $\Delta N^{-1}$ and $N N^{-1}$ components as well, 
and their strengths vary with the momentum (and density). 
All these effects are properly taken into account 
in the formalism of ref.\ \cite{simAP}, 
and the proper partial widths 
are obtained from eq.\ (\ref{eq_GnuReal}).

The general behavior of the partial widths to the pionic modes 
are that for low invariant $\Delta$ mass 
(keeping $p_\Delta$=$0$ for the moment) 
the decay to the lower pionic mode 
resembles the decay to a free pion. 
As the invariant $\Delta$ mass increases 
(and the pionic momentum correspondingly),
the partial width becomes smaller than the free width 
because the pion amplitude on the lower mode decreases. 
For further increasing invariant mass 
the partial width to the lower mode starts to decrease 
and becomes very small. 
This is because the pion component vanishes 
and the collectivity disappears when the mode enters 
the band of non-collective $\Delta N^{-1}$ modes. 
Note that although the decay width 
to the band of $\Delta N^{-1}$ modes can be quite substantial, 
the decay to a single $\Delta N^{-1}$ mode is small 
and vanishes in the continuum limit (box-size $\rightarrow \infty$). 
Thus very few lower pionic modes ($\tilde{\pi}_1$) 
with large momentum 
(in the local medium frame)
will be created, 
and above a certain momentum (depending on the density) 
no $\tilde{\pi}_1$ will be created at all. 
Instead,
the partial decay width to the upper pionic mode 
is substantial for large invariant mass,
where this mode takes over the role of the free pion.

Another feature for the $\Delta$ width in the medium  
is that it depends explicitly on the $\Delta$ momentum in the medium 
(in addition to the $\Delta$ energy), 
while in vacuum the relativistic invariant mass 
$m=\sqrt{E_\Delta^2 + p_\Delta^2 c^2}/c^2$ 
suffices to determine the width. 
For the total width this momentum dependence is rather weak 
in comparison with the dependence on the invariant mass 
and may be neglected in a first approximation. 
However, for the partial widths this is a poor approximation, 
especially for the decay to the lower mode 
(where the width becomes small for large pionic momenta 
in the medium frame).
Thus a different treatment for the $\Delta$ decay is needed, 
as compared to the standard transport prescription 
where the decay takes place isotropically in the $\Delta$ rest frame, 
with the $\Delta$ decay width determined by the invariant mass.
To this end we calculate and store in a large table 
the partial $\Delta$ decay widths to the two pionic modes, 
for a mass-momentum grid in the medium frame. 
In addition, these decay widths are obtained 
for a number of different densities, $\rho_i$. 
The $\Delta$ then decays in the medium frame 
according to linear interpolation in the table of decay widths.

\begin{figure}	%.............................................................
\centerline{\psfig{file=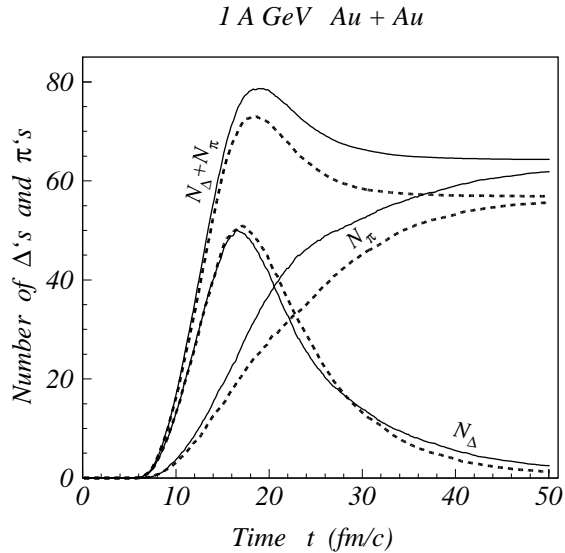,height=8.0cm,angle=0}}
\caption{Time evolution of pions and $\Delta$ resonances.
         Solid curves represent simulations 
         with in-medium properties included, 
         while dashed curves correspond to standard BUU simulations.}
\label{fig_PiAndDelta}
\end{figure}	%.............................................................

% =====================================================================
\section{Results and discussion}
\label{sec_Res}
% =====================================================================
In this section we present results from BUU simulations, 
containing in-medium effects as discussed 
in section \ref{sec_SIMtoTM}. 
The purpose of these simulations is to elucidate 
the role and effects of the included in-medium properties.
We therefore compare our results to standard BUU simulations. 
Further, to make the treatment as simple as possible 
(and the signals as clean as possible), 
we have excluded some processes 
that may be of equal importance when comparing with experimental results. 
Such processes include direct pion production, 
$N^*$ resonances,
and impact parameter averaging,
as well as other in-medium properties not treated here 
(though discussed to some extent).

We have performed simulations for central collisions of
two different symmetric systems, 
one light, $^{40}$Ca+$^{40}$Ca, 
and one heavy $^{197}$Au+$^{197}$Au, 
each at two different energies, 500 and 1000~$A$~GeV. 
All simulations were performed with 250 test particles per nucleon,
utilizing a mean field
$U(\rho) = -0.218 \, (\rho/\rho_0) + 0.164 \, (\rho/\rho_0)^{4/3}$ (GeV).
Figure \ref{fig_PiAndDelta} presents 
the time evolution of the pions and $\Delta$'s 
for the Au+Au reaction at 1~$A$~GeV.
The net effect in the total number of pions 
escaping the collision as free 
is that somewhat more pions are produced in the medium-modified simulations.
The difference is about 10\% 
for both systems and bombarding energies.

For the medium-modified simulations 
we find that at the higher energy about 95\% 
of the emitted pions originate from the lower pionic mode, 
while the lower mode is even more dominant 
at the lower bombarding energy,
with a contribution of about 99\%. 
For the Au system we find that 2-5\% 
of the pionic modes 
are realized as (unphysical) $\de N^{-1}$ states, 
while for the Ca system the $\de N^{-1}$ realization 
amounts to 7-9\%. 
For the upper pionic mode, though, the situation is worse
since about half of these modes end up with the $\de N^{-1}$ realization. 
However, since their total contribution is very small, 
also the total error is small.

\begin{figure}	%.............................................................
\centerline{\psfig{file=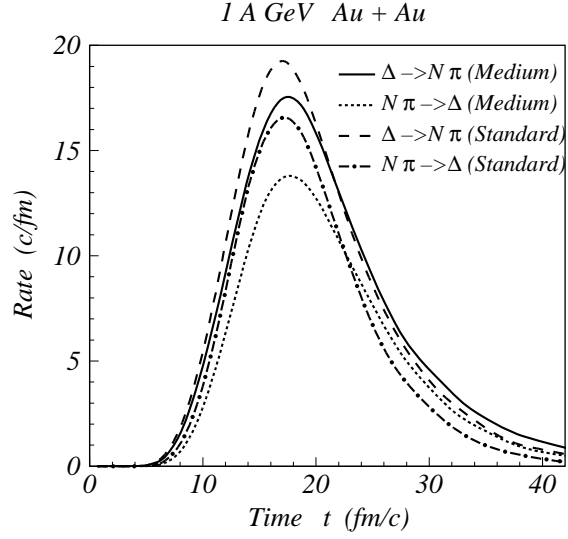,height=8.0cm,angle=0}}
\caption{The time dependence of the rates for
$\Delta$ decay and pion reabsorption.}
\label{fig_DecayAndReab}
\end{figure}	%.............................................................

Figure \ref{fig_DecayAndReab} displays the number of $\Delta$ decays 
($\Delta \rightarrow N + \tilde{\pi}$) 
and pion reabsorption processes ($N + \tilde{\pi} \rightarrow \Delta$)
per unit time,
for the Au+Au reaction at 1~$A$~GeV.
Comparing the rate of decay processes 
($\de \rightarrow N + \tilde{\pi}$) 
for both the heavy system and the light system,
we find that they look quite similar in the two types of transport simulation.
In both systems there is a tendency to a few per cent less decay 
during the dense phase in the medium-modified simulation 
at the high bombarding energy.

\begin{figure}	%.............................................................
\centerline{\psfig{file=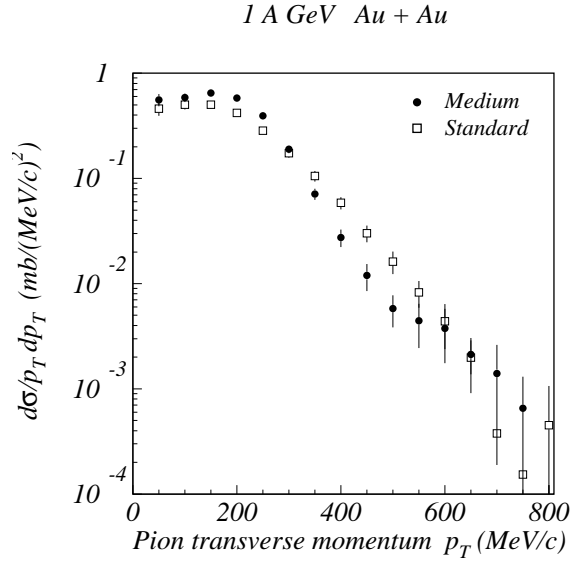,height=8.0cm,angle=0}}
\caption{Transverse momentum spectrum for neutral pions
         in the central rapidity interval $-0.16 < y_{\rm cm} < 0.16$.
         Filled circles represents simulations 
         with medium properties included, 
         while open squares correspond to the standard simulations.
         The error bars represent the statistical errors.}
\label{fig_pT}
\end{figure}	%.............................................................

However,
the numbers of reabsorbed quasi-pions 
($N+\tilde{\pi} \rightarrow \de$) 
differ for two treatments:
When the medium properties are included,
there is 5-20\% 
less reabsorption in the dense phase, 
while the reabsorption is slightly larger 
at late times when the system is dilute. 
The effect is largest for the Au system at 1~$A$~GeV 
and smallest for the Ca system at 500~$A$~MeV.
Inspecting the time and density distributions of the reabsorbed pions,
we also find that the absorption 
is somewhat delayed in the medium-modified simulations 
as compared to the standard simulations. 
Furthermore, 
when the medium effects are incorporated,
the pionic population (and consequently the absorption) 
is smaller in the dense regions, 
while the reabsorption is enhanced at low densities 
($\rho < 0.5 \rho_0$).

In our work, the reabsorption is  
modified only by the in-medium pionic dispersion relation, 
$\hbar \tilde{\omega}$, 
which is used to determine the energy of the produced $\Delta$.
As discussed in ref.\ \cite{simAP}, 
the full medium corrections 
to the $\tilde{\pi} + N \rightarrow \Delta$ cross section 
should also contain a modification 
of the interaction factor due to the decrease 
of the pion component on the pionic mode 
in favor of the $N N^{-1}$ and $\Delta N^{-1}$ components, 
as well as the total in-medium $\Delta$ decay width 
in the Breit-Wigner factor.
Since the reabsorption cross section 
has a rather sharp resonance peak, 
relatively moderate changes in the $\Delta$ energy 
can produce rather large effects 
in the number of absorbed pions.
The reabsorption cross section 
is expected to be reduced 
when taking into account 
the full in-medium effects. 
Therefore we expect the reabsorption effect 
to be even stronger in a simulation 
taking into account also these in-medium effects.
However, the net reabsorption depends strongly 
on the energy distribution of the pionic modes 
which may affect such an expectation.

Figure \ref{fig_pT} presents 
the transverse momentum spectrum 
$d\sigma/{p_T}{dp_T}$ for the Au+Au reaction at 1~$A$~GeV.
Only zero impact parameter has been used 
and the cross sections have been obtained 
by assuming that all collisions with an impact parameter up to 
$b_{\rm max}$=$2 \, r_0 A^{1/3}_{\rm Au}$ contribute equally. 
Thus this cross section represents an upper estimate.
The effect seen in the simulations incorporating the in-medium properties, 
as compared to the standard simulations, 
is a modest but significant enhancement at low transverse momenta 
and a reduction at higher momenta,
corresponding to a reduction of the effective transverse temperature.
This effect is seen in all four simulations, 
but is naturally most pronounced 
at the higher energy and the heavier system. 
This effect is due to the lower pionic mode 
and the partial width for $\Delta$ decay to this mode. 
As discussed in sect.\ \ref{sec_SIMtoTM},
very few $\tilde{\pi}_1$ modes are created 
at high momentum in the local medium frame.
The upper mode $\tilde{\pi}_2$ 
which may be created at arbitrary high momentum 
comes higher in $\Delta$ energy 
and thus not many of these modes are created either. 
So the net effect
is an enhancement of low energy pions 
and a reduction of high energy pions.

We finally wish to emphasize that
the spectra presented in fig.\ \ref{fig_pT} 
are not entirely suitable for quantitative comparison
with experimental spectra,
partly because of the simple impact-parameter averaging employed,
and partly because the incorporation of higher nucleon resonances 
and additional medium effects might be of importance. 
Nevertheless,
we expect that the enhancement of low-energy pions
would survive such a more complete simulation.

In summary,
we have performed exploratory simulations
with an existing transport code into which we have incorporated
in-medium spin-isospin properties calculated in a microscopic model
that is more consistent than those employed in earlier studies.
The effects on the dynamical evolution of pions and $\Delta$ isobars
have been elucidated and, in particular,
it appears that the medium-modified treatment leads to a lowering
of the effective temperature of the transverse pion spectra.
Our results suggest that at the quantitative level
it is important to take account of these in-medium effects
in simulations of nuclear collisions at intermediate energies.

% ----------------------------------------------------------------------
% \acknowledgements
% ----------------------------------------------------------------------
\newpage
Stimulating discussions 
with Wolfgang Bauer and Volker Koch 
are acknowledged.
Wolfgang Bauer is also acknowledged 
for making his BUU code available to us.
J.H. would like to thank the Nuclear Science Division at
Lawrence Berkeley National Laboratory for support and hospitality,
and J.R. wishes to thank the ECT$^*$ in Trento for support and hospitality. 
This work was supported
by the Training and Mobility through Research (TMR) programme
of the European Community under contract ERBFMBICT950086
and by the Director, Office of Energy Research, 
Office of High Energy and Nuclear Physics, 
Nuclear Physics Division of the U.S. Department of Energy 
(Contract DE-AC03-76SF00098).
% --------------------------------------------------------------------

% ====================================================================
%                             References                             =
% ====================================================================

% \begin{references}

% \end{references}

\end{document}